\documentclass[a4paper]{article}
\usepackage{Odyssey2018}
\usepackage{epsfig,amssymb,amsmath,subfigure,epsfig,siunitx,color,multirow,graphicx,bm,textcomp,enumitem}
\usepackage{bigstrut}% Bigstrut package, forces specific row heights in tables to be larger

\newcommand{\argmax}{\arg\!\max}

\ninept
\title{A Regression Model of Recurrent Deep Neural Networks for Noise Robust Estimation of the Fundamental Frequency Contour of Speech}

\name{Akihiro Kato, Tomi Kinnunen}

\address{School of Computing  \\
University of Eastern Finland \\
{\small \tt akihiro.kato@uef.fi, tomi.kinnunen@uef.fi} }
\begin{document}
\maketitle

\begin{abstract}
The fundamental frequency ($F0$) contour of speech is a key aspect to represent speech prosody that finds use in speech and spoken language analysis such as voice conversion and speech synthesis as well as speaker and language identification. 

This work proposes new methods to estimate the $F0$ contour of speech using deep neural networks (DNNs) and recurrent neural networks (RNNs). They are trained using supervised learning with the ground truth of $F0$ contours. The latest prior research addresses this problem first as a frame-by-frame-classification problem followed by sequence tracking using deep neural network hidden Markov model (DNN-HMM) hybrid architecture. This study, however, tackles the problem as a regression problem instead, in order to obtain $F0$ contours with higher frequency resolution from clean and noisy speech.

Experiments using \emph{PTDB-TUG} corpus contaminated with additive noise (\emph{NOISEX-92}) show the proposed method improves gross pitch error (GPE) by more than 25 \% at signal-to-noise ratios (SNRs) between -10 dB and +10 dB as compared with one of the most noise-robust $F0$ trackers, PEFAC. Furthermore, the performance on fine pitch error (FPE) is improved by approximately 20 \% against a state-of-the-art DNN-HMM-based approach.
\end{abstract}

\section{Introduction\label{sec:intro}}
The \emph{fundamental frequency} ($F0$) represents the lowest frequency in a quasi-periodic signal. In human speech production $F0$ is determined by the movement of the vocal chords and the contour of $F0$s represents important aspects of prosody. Therefore, $F0$ is one of the key features of speech and is used in many applications including voice conversion \cite{mohammadi17}, speaker and language identification \cite{torres17, nandi17}, prosody analysis \cite{godoy17}, speech coding \cite{rajendran17}, speech synthesis \cite{wang17-2} and speech enhancement \cite{kato14, kato16}.

Over the past decades, a variety of approaches to $F0$ estimation have been proposed. Specifically, Robust Algorithm for Pitch Tracking (RAPT) \cite{talkin95} and YIN \cite{kawahara02} that exploit autocorrelation of a time-domain signal are among the best methods to estimate $F0$ and have been widely used in many applications \cite{pirker11}. However, it is well known that these methods do not produce satisfactory results under noisy conditions \cite{wang14}. Several alternative robust frequency- and cepstral-domain $F0$ estimators have been developed. For instance, Pitch Estimation Filter with Amplitude Compression (PEFAC) \cite{gonzalez14} has high performance in noisy conditions. It analyses noisy signals in the log-frequency domain with a matched filter and the universal long-term average speech spectrum. Nonetheless, it is still challenging to achieve sufficient $F0$ estimation accuracy at low signal-to-noise ratios (SNRs) such as 0 dB and below.

In addition to the instantaneous signal processing methods mentioned above, various machine learning approaches that utilise generative models, %based on joint probability densities 
such as a Gaussian mixture model (GMM) and hidden Markov models (HMMs) \cite{milner07, le07, sha04, jin11}, have been developed along with particle filters \cite{zhang16-1, hajimolahoseini16} to address the challenge related to severe noisy conditions. In this context, models based on deep neural networks (DNNs) have shown promising achievement in tackling the problem \cite{wang14, han14-1, han14-2} because of the explicit capability of DNNs for complex pattern mapping as a discriminative model.

For classification problems, %it is known that 
discriminative models can outperform generative models when trained from large enough quantity of data \cite{ng02}. DNNs derive a discriminative model to represent arbitrarily complex functions as long as they consist of large number of units in their hidden layers. Consequently, they enable statistical models to deal with higher dimensional input features having stronger correlation than the preceding generative models. Thus, they have been successfully applied to various speech applications showing great advantages in performance over the existing statistical models \cite{yu11, hinton12, zen13, zhang16-2, xu14}. Furthermore, recurrent neural networks (RNNs), in which each unit has a connection pointing backward from its output to the input, might be more suitable for time series signals of speech to track temporal dynamics.

In fact, the latest research have proposed DNN and RNN-based models for $F0$ estimation, showing improvement in noise robustness as compared to the conventional algorithms \cite{wang14, han14-1, han14-2, verma16, liu17, wang17-1}. These state-of-the-art $F0$ estimators, however, still have a problem to be solved: they first employ DNNs or RNNs to form a frame-by-frame \emph{classification} model to decide a frequency state corresponding to \emph{quantised} frequency, followed by frame-by-frame tracking to optimize the most likely state sequence. This is achieved by utilising a hybrid deep neural network hidden Markov model (DNN-HMM) architecture \cite{dahl12} that has a successful history, for instance, in automatic speech recognition (ASR) \cite{chan16} and text-to-speech (TTS) \cite{wang17-2}. Even if it is convenient to treat $F0$ tracking as a classification task analogous to speech recognition, the resulting estimated $F0$ contour has a limited frequency resolution determined by the number of frequency states. This is a potential draw-back in applications that require high-precision $F0$ values, such as voice conversion, or micro-prosody analysis for speaker and language characterisation.

%potentially gives a drawback to $F0$-sensitive applications such as prosody analysis and language identification and gives motivation to our work.

To sum up our contribution, we aim at improving the state-of-the-art DNN and RNN-based approaches to $F0$ estimation in terms of increased tracking precision \emph{and} noise robustness, by treating the problem as a \emph{regression} task instead of the DNN-HMM-based classification tasks reviewed above. Even if our work is not the first work to address $F0$ tracking as a regression task where $F0$ values are predicted from other speech representations \cite{milner07, le07, sha04, jin11}, we do improve over the latest deep learning approaches. For maximum applicability of our results, we treat the problem in a speaker-independent manner, and study the sensitivity of the results under both unknown and known noise conditions.

%\section{Framework of a DNN (RNN)-based F0 estimator\label{sec:framework}}
\section{General framework\label{sec:framework}}

\begin{figure}[htbp]
	\begin{center}
		\includegraphics[scale=0.38]{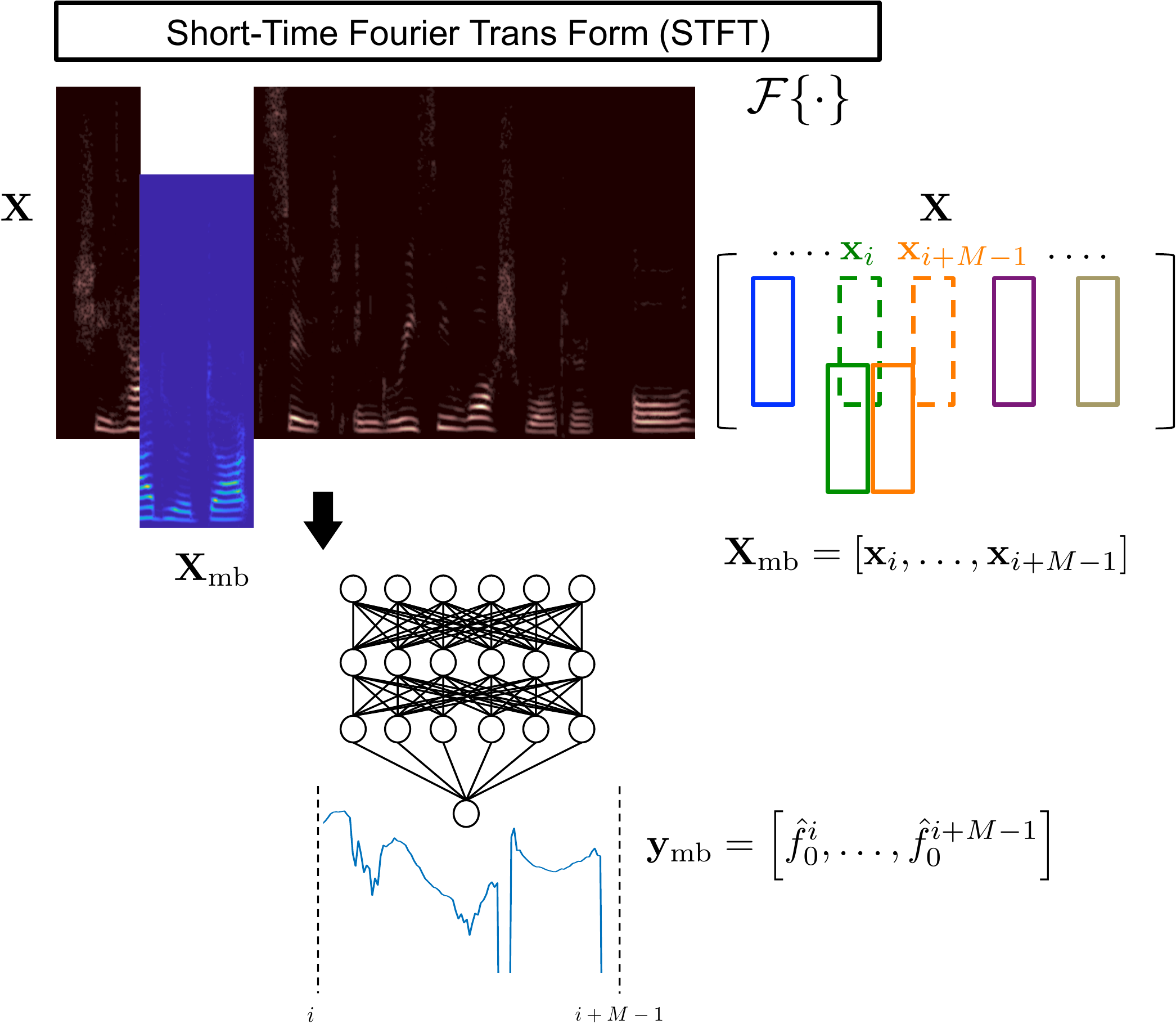}
		\caption{\it Overview of the DNN-based F0 estimation showing $M$ frames are extracted from sequence of magnitude spectra, ${\bf X}$, into mini-batch, ${\bf X}_{\text{mb}}$, and then mapped onto $F0$ contour, ${\bf y}_{\text{mb}}$.}
		\label{fig:overview}
	\end{center}
\end{figure}

Before presenting our proposal in Section \ref{sec:model}, we first review a general framework of a DNN-based approach to $F0$ contour estimation. A speech signal is first converted to a sequence of magnitude spectra by short-time Fourier transform (STFT). A DNN-based $F0$ estimation model trained by supervised learning then maps the spectral information to the fundamental frequency contour, as illustrated in Fig. \ref{fig:overview}.

Discrete time-domain speech signal, $s(n)$, is divided into $I$ frames, $s_0(m), s_1(m), \dots, s_{I-1}(m)$, where $m$ denotes time sequence in a window function, and then transformed to the frequency domain by STFT to derive a sequence of magnitude spectra, ${\bf X}$ as
\begin{eqnarray}
	{\bf X} &=& \left[{\bf x}_0, {\bf x}_1, \dots, {\bf x}_{I-1}\right]\label{eq:x_seq}\\
	{\bf x}_i &=& \left[|x_i(1)|, |x_i(2)|, \dots, |x_i(K)|\right]^{\top}\\
	x_i(k) &=& \mathcal{F}\left\{s_i(m)\right\}
\end{eqnarray}
where $\mathcal{F}$\{$\cdot$\} denotes the discrete \emph{Fourier transform} (DFT) and $K$ represents the number of DFT bins between 0 Hz and the Nyquist frequency of $s(n)$. 

The input to the DNN is a subset of ${\bf X}$ which defines mini-batch input, ${\bf X}_{mb}$, as
\begin{equation}
	{\bf X}_{\text{mb}} = \left[{\bf x}_{\mu0}, {\bf x}_{\mu1}, \dots, {\bf x}_{\mu(M-1)}\right]
\end{equation}
where
\begin{equation}
	\{\mu0, \mu1, \dots, \mu(M-1)\} \subset \{0, 1, \dots, I-1\}
\end{equation}
and $M$ represents the mini-batch size.

Since the DNN is trained by supervised learning, the DNN maps the inputs onto their target $F0$ values or states. Consequently, estimates of the $F0$s or pitch candidates are finally derived by the DNN.
\begin{equation}
	{\bf y}_{\text{mb}} = \left[\hat{f}_0^{\mu0}, \hat{f}_0^{\mu1}, \dots, \hat{f}_0^{\mu(M-1)}\right]^{\top}
\end{equation}
where $\hat{f}_0^i$ is the estimate of $F0$ at the $i$-th frame.

To capture the characteristics of the temporal dynamics in speech in addition to the static feature within a frame, RNNs can be applied to the $F0$ estimation model instead of DNNs. The details of neural network models for the preceding framework are discussed in Section \ref{sec:model}.

\section{DNN (RNN)-based regression models for F0 estimation\label{sec:model}}
This section first discusses the DNN model for the proposed method to estimate the $F0$ contour of speech in Section \ref{sec:dnn} followed by the RNN model for the proposed method in Section \ref{sec:rnn}.

\subsection{DNN model\label{sec:dnn}}
Since the neighbouring frames of the $i$-th frame may contain useful information to estimate $f_0^i$ \cite{han14-1}, 
the input mini-batch of the DNN model includes augmented vector, ${\bf x}^i$, which comprises ${\bf x}_i$ and its context as
\begin{equation}
	{\bf x}^i = \left[{\bf x}_{i-p}^{\top}, \dots, {\bf x}_{i-1}^{\top}, {\bf x}_i^{\top}, {\bf x}_{i+1}^{\top}, \dots, {\bf x}_{i+p}^{\top}\right]^{\top}
\end{equation}
where $p$ denotes the number of the context frames which are added to the both side of ${\bf x}_i$. Therefore, the input of the DNN is illustrated as
\begin{eqnarray}
	{\bf X}_{\text{mb}} &=& \left[{\bf x}^{\mu0}, \dots, {\bf x}^{\mu(M-1)}\right]\\
	&=& \left[
	\begin{array}{ccc}
		{\bf x}_{\mu0-p} & \dots & {\bf x}_{\mu(M-1)-p}\\
		\vdots & \ddots & \vdots\\
		{\bf x}_{\mu0} & \dots & {\bf x}_{\mu(M-1)}\\
		\vdots & \ddots & \vdots\\
		{\bf x}_{\mu0+p} & \dots & {\bf x}_{\mu(M-1)+p}
	\end{array}
	\right]
\end{eqnarray}
where the frame indexes below zero are set to zero while the frame indexes over $I-1$ are set to $(I-1)$ because the range of the frame indexes are determined by Equation (\ref{eq:x_seq}).

For mini-batch input, ${\bf X}_{\text{mb}}$, output of the $l$-th layer of the DNN, ${\bm \Theta}^l$ is derived as
\begin{eqnarray}
	{\bm \Theta}^l &=& g\left({\bf W}^l{\bm \Phi}^l\right)\\
	&=& \left[{\bm \theta}_0^l, {\bm \theta}_1^l, \dots, {\bm \theta}_{M-1}^l\right]\\
	&=& \left[
	\begin{array}{cccc}
		\theta_{10}^l & \theta_{11}^l & \dots & \theta_{1(M-1)}^l\\
		\theta_{20}^l & \theta_{21}^l & \dots & \theta_{2(M-1)}^l\\
		\vdots & \vdots & \ddots & \vdots\\
		\theta_{q_l0}^l & \theta_{q_l1}^l & \dots & \theta_{q_l(M-1)}^l
	\end{array}
	\right]\\	
	{\bf W}^l &=& \left[
	\begin{array}{cccc}
		w_{10}^l & w_{11}^l & \dots & w_{1q_{l-1}}^l\\
		w_{20}^l & w_{21}^l & \dots & w_{2q_{l-1}}^l\\
		\vdots & \vdots & \ddots & \vdots\\
		w_{q_l0}^l & w_{q_l1}^l & \dots & w_{q_lq_{l-1}}^l
	\end{array}
	\right] \label{eq:weight}\\
	{\bf \Phi}^l &=& \left[
	\begin{array}{cccc}
		1 & 1 & \dots & 1\\
		{\bm \theta}_0^{l-1} & {\bm \theta}_1^{l-1} & \dots & {\bm \theta}_{M-1}^{l-1}\\
	\end{array}
	\right]\\
	{\bm \theta}_\lambda^0 &=& {\bf x}^{\mu\lambda}
\end{eqnarray}
where $q_l$ denotes the number of units excluding the bias unit in the $l$-th layer, $w_{jk}^l$ is the weight between unit, $k$, in the ($l-1$)-th layer and unit, $j$, in the $l$-th layer. Lastly, $g(\cdot)$ represents an activation function.

In the context of $F0$ estimation, it is common to sort out the problem as a classification task by applying the softmax function to the output layer consisting of $U$ units in order to exploit a DNN-HMM framework \cite{wang14, han14-1, han14-2, verma16}. In such cases, frequency states, $s_u \in \{s_0, s_1, \dots, s_{U-1}\}$, representing \emph{quantised frequency} are determined. Outputs from the DNN model give \textit{a posteriori} probabilities of each frequency state, $P(s_u|{\bf x^i}),\ \forall u = 0, 1, \dots, U-1$, at the $i$-th frame. Therefore, estimate of the $F0$ contour, $\hat{\bf f}'_0$, is obtained by tracking the most likely frequency state at each frames. 

Since \textit{a priori} probability $P(s_u)$ is computed during training, where transition probabilities from $s_u$ to $s_v,\ \gamma_{uv},\ \forall u, v = 0, 1, \dots, U-1$, are also computed, Bayes' theorem derives $P({\bf x}^i|s_u)$ as
\begin{equation}
	P\left({\bf x}^i\mid s_u\right) \propto \frac{P\left(s_u|{\bf x}^i\right)}{P\left(s_u\right)}
\end{equation}
Hence, the Viterbi algorithm optimises $\hat{\bf f}'_0$ as
\begin{eqnarray}
	\hat{\bf f}_0 &=& \argmax_{\hat{\bf f}'_0}{P\left(\hat{\bf f}'_0\mid \gamma_{uv}, P({\bf x}^i\mid s_u), {\bf X}\right)}\\
	&\ & ~~~~~~~~~~~~~~~~~\text{for } \forall u, v = 0, 1, \dots, U-1\nonumber\\
	&\ & ~~~~~~~~~~~~~~~~~\text{for } \forall i = 0, 1, \dots, I-1\nonumber
\end{eqnarray}
where
\begin{equation}
	{\bf X} = \left[{\bf x}^0, {\bf x}^1, \dots, {\bf x}^{I-1}\right]
\end{equation}
This is referred to as DNN-HMM hybrid architecture being successfully applied to many applications as mentioned in Section \ref{sec:intro}. However, $\hat{\bf f}_0$, still comprises quantised values.

Alternatively, in the proposed method the $F0$ estimation model with $L$ layer DNNs, i.e.\ DNNs consisting of ($L-1$) hidden layers and an output layer, sets $q_L$ equal to 1 and applies the identity function to the output layer. Consequently, the DNNs map the input onto the target value directly as a regression model as follows.
\begin{eqnarray}
	{\bm \Theta}^L &=& g\left({\bf W}^L{\bm \Phi}^L\right)\\
	&=& \left[w_{10}^L \dots w_{1q_{L-1}}^L\right]
	\left[
	\begin{array}{ccc}
		1 & \dots & 1\\
		{\bm \theta}_0^{L-1} & \dots & {\bm \theta}_{M-1}^{L-1}
	\end{array}
	\right]~~~~~\\
	&=& \left[\hat{f}_0^{\mu0} \dots \hat{f}_0^{\mu(M-1)}\right]\\
	&=& \left({\bf y}_{\text{mb}}\right)^{\top}
\end{eqnarray}
where $g(\cdot)$ is an identity function.

In the offline training process, ${\bm W}^l$ is optimised in advance by mini-batch gradient descent with the backpropagation algorithm \cite{rumelhart85} to minimise the MSE between ${\bm \Theta}^L$ and the ground truth of the $F0$ contour.

\subsection{RNN model\label{sec:rnn}}
Units in RNN layers have connections to send their outputs back to their own inputs in addition to the feedforward connections. Therefore, an RNN layer receives its own output at the previous time sequence as well as the current time sequence input from the previous layer. This behaviour of RNN layers interpreted as memory cells is suitable to analyse temporal dynamics in speech signals. Therefore, each instance in ${\bf X}_{\text{mb}}$ of RNNs includes only a frame of one time sequence, unlike instances in ${\bf X}_{\text{mb}}$ of DNNs concatenated with the neighbouring frames, and its temporal context are analysed sequence-to-sequence by exploiting the memory cells of RNNs. Accordingly, a time sequence of the RNN inputs is represented as follows.
\begin{eqnarray}
	{\bf X}_{\text{mb}}^{0} &=& \left[{\bf x}_{\mu0-p}, \dots, {\bf x}_{\mu(M-1)-p}\right]\nonumber\\
	&\vdots&\nonumber\\
	{\bf X}_{\text{mb}}^{p-1} &=& \left[{\bf x}_{\mu0-1}, \dots, {\bf x}_{\mu(M-1)-1}\right]\nonumber\\
	{\bf X}_{\text{mb}}^{p} &=& \left[{\bf x}_{\mu0}, \dots, {\bf x}_{\mu(M-1)}\right]\nonumber\\
	{\bf X}_{\text{mb}}^{p+1} &=& \left[{\bf x}_{\mu0+1}, \dots, {\bf x}_{\mu(M-1)+1}\right]\nonumber\\
	&\vdots&\nonumber\\
	{\bf X}_{\text{mb}}^{2p} &=& \left[{\bf x}_{\mu0+p}, \dots, {\bf x}_{\mu(M-1)+p}\right]
\end{eqnarray}
where 
\begin{equation}
	\left\{\mu0, \mu1, \dots, \mu(M-1)\right\} \subset \left\{0, 1, \dots, I-1\right\}
\end{equation}
${\bf X}_{\text{mb}}^n$ represents the mini-batch of the RNN input at time sequence, $n$. $M$ and $I$ denote the mini-batch size and the total number of frames in the dataset respectively while $p$ is the period to analyse temporal context, i.e.\ $p$ for both of the past and future makes $2p+1$ time-sequence analysis in totall.

\begin{figure}[htbp]
	\begin{center}
		\includegraphics[scale=0.39]{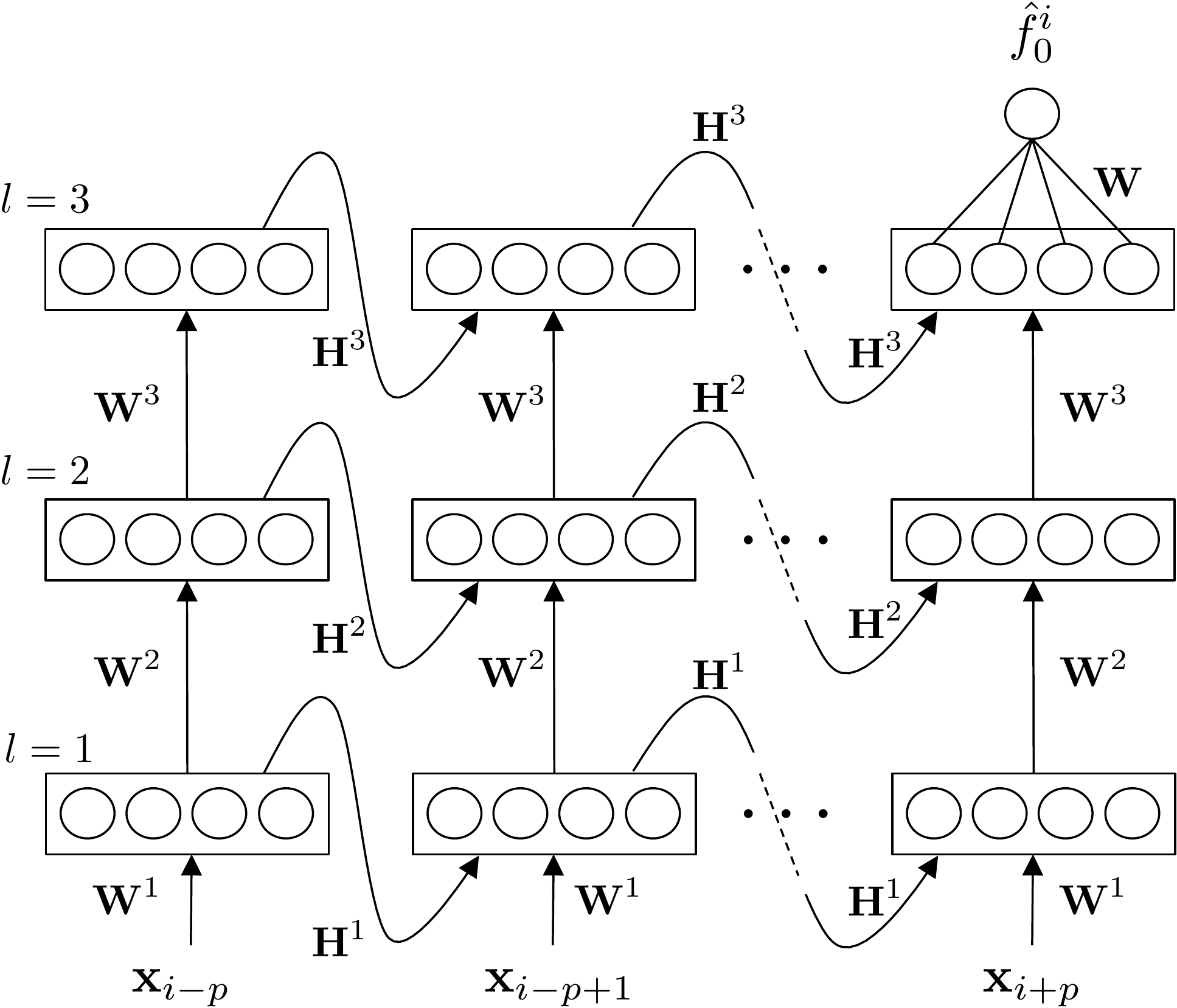}
		\caption{\it An unrolled diagram showing the RNN-based model of the proposed method. The model is formed taking an encoder structure.}
		\label{fig:rnn_form}
	\end{center}
\end{figure}

The RNN model in the proposed method takes a form of \emph{encoder} structure as illustrated in Figure \ref{fig:rnn_form}. 
Output of RNN layer, $l$, at time sequence, $n$ $(n = 0, 1, \dots, 2p)$, ${\bm \theta}_n^l$, is derived as follows with respect to one instance, ${\bf x}_i$, in a mini-batch.
\begin{eqnarray}
	{\bm \theta}_n^l &=& g\left({\bf W}^l{\bm \phi}_n^l + {\bf H}^l{\bm \phi}_{n-1}^{l+1}\right)\\
	{\bm \phi}_n^l &=& \left[1, ({\bm \theta}_n^{l-1})^{\top}\right]^{\top}\\
	{\bm \theta}_n^0 &=& \left[1, ({\bf x}_{i-p+n})^{\top}\right]^{\top}
\end{eqnarray}
where ${\bf W}^l$ is the weight matrix from the output of layer $l-1$ to the input of layer $l$ (feedforward) while ${\bf H}^l$ denotes the weight matrix from the output of layer $l$ to the input of layer $l$ (feedback). The form of ${\bf W}^l$ and ${\bf H}^l$ is same as weight matrices of DNNs shown in Equation (\ref{eq:weight}).

Only the last time sequence has the output layer consisting of a unit connected with the previous RNN layer with feedforward weight matrix, ${\bf W}$, to output the estimate of $F0$ at frame $i$.
\begin{eqnarray}
	y &=& g\left({\bf W}{\bm \phi}_{2p}^L\right)\\
	&=& \hat{f}_0^i
\end{eqnarray}
where $g(\cdot)$ is the identity function and $L$ denotes the number of RNN layers. Therefore, this algorithm is equivalent to an encoder, in which a sequence of observation, $[{\bf x}_{i-p}, \dots, {\bf x}_{i-1}, {\bf x}_i, {\bf x}_{i+1}, \dots, {\bf x}_{i+p}]$, is encoded to $\hat{f}_0^i$.

\section{Experiments\label{sec:result}}

In our experiments we address both accuracy and noise robustness of the proposed methods and compare them with \emph{RAPT} \cite{talkin95}, \emph{YIN} \cite{kawahara02}, \emph{PEFAC} \cite{gonzalez14} and a state-of-the-art DNN-HMM-based approach (\emph{DNN-HMM}) \cite{han14-2} as representatives of existing methods. \cite{han14-2} also has reported that there was no significant difference in performance between DNN-HMM and their RNN-based classification approach. Therefore, we selected DNN-HMM approach for the competition of our approach.

\subsection{Datasets (PTDB-TUG corpus)}
For the experiments, we adopt PTDB-TUG corpus \cite{pirker11}. Since the proposed methods and DNN-HMM require a training and cross-validation (CV) datasets for offline training, the training set constitutes of 2640 utterances spoken by 16 speakers (8 males / 8 females), 165 utterances from each. The CV dataset consists of other 576 utterances spoken by the same 16 speakers, i.e.\ 36 utterances each. For the test dataset, we use 944 utterances spoken by 4 unknown speakers (two males / two females) that are not in the training and CV datasets (\emph{unknown} speakers), i.e.\ 236 utterances each, are contained as well as other 560 utterances spoken by the 16 speakers (\emph{known} speakers) in order to make a \emph{speaker independent} (SI) test set. Table \ref{tab:corpus} summarises the data allocation to each dataset.
\begin{table}[htbp]
	\begin{center}
		\caption{\it Data allocation from PTDB-TUG to the datasets. Utts, Spkrs, F and M are abbreviations for Utterances, Speakers, Females and Males respectively.}
		\label{tab:corpus}
		\begin{tabular}{l|c|c|r}
			Subset & Speakers & Utts (/ Spkr) & Duration\\
			\hline
			Training & 8 F + 8 M & 2,640 (165) & 307 min\\
			CV & 8 F + 8 M & 576 (36) & 67 min\\
			\hline
			\multirow{2}{*}{Test} & \footnotesize{8 F + 8 M (Known)} & 560 (35) & 61 min\\
			& \footnotesize{2 F + 2 M (Unknown)} & 944 (236) & 104 min
		\end{tabular}
	\end{center}
\end{table}

The PTDB-TUG corpus contains ground truths $F0$ contours of each utterances, obtained from laryngograph signals recorded in a clean condition to which a Kaiser filter and RAPT are applied. They are used in the following experiments as the ground truth.

\subsection{Noisy conditions (NOISEX-92)}
Speech in each dataset is sampled at 16kHz and the sampled signals in the training and CV datasets are contaminated with six types of additive noise at five levels of SNR while eight types of additive noise at five levels of SNR are added to the test dataset. The eight types of noise for the test dataset are referred to as Babble, F16, Factory1, Leopard, Machinegun, Pink, Volvo and White in NOISEX-92 \cite{varga93}. Factory1 and Pink are not applied to the training and CV datasets so that these two types of noise play a role of unknown noise for the proposed methods and DNN-HMM at tests. All the utterances in the datasets make noisy speech with each noise type at SNRs of -10, -5, 0, 5 and 10 dB. Consequently, the training dataset amounts 81,840 utterances (9,517 min), i.e.\ 2,640 $\times$ (6 noise $\times$ 5 level + 1 clean), and the CV dataset becomes 17,856 utterances (2,077 min) while the test dataset amount 60,160 utterances (6,600 min), i.e.\ (560 + 944) $\times$ 8 noise $\times$ 5 level, in total. Table \ref{tab:noise} summarises the noise types used for training and test sets.

\begin{table}[htbp]
	\begin{center}
		\caption{\it The summary of additive noise used in experiments.}
		\label{tab:noise}
		\begin{tabular}{l|c|c|c}
			Type (NOISEX-92) & Training & Test & Stationarity\\
			\hline
			Clean & Yes & No & -\\
			Babble & Yes & Yes & Low\\
			F16 & Yes & Yes & High\\
			\bf Factory1 & \bf No & \bf Yes & \bf Low\\
			Leopard & Yes & Yes & Low\\
			Machinegun & Yes & Yes & Low\\
			\bf Pink & \bf No & \bf Yes & \bf High\\
			Volvo & Yes & Yes & High\\
			White & Yes & Yes & High
		\end{tabular}
	\end{center}
\end{table}

\subsection{Training and Test settings}
The speech signals in the datasets are framed into 25 ms frames at 5 ms intervals and then the first 400 frames and 200 frames at the tail in each utterance are removed to reduce non-speech frames. For frequency-domain analysis of the proposed methods, STFT is applied with 1024-point FFT in order to obtain time-frequency domain power spectral density (PSD) and the first 513 bins in the frequency-domain, i.e.\ $0 \le \omega \le \pi$, at each frame are used for the mini-batch analysis. Procedures for feature extraction and $F0$ quantisation for DNN-HMM follow \cite{han14-2}.

RAPT, YIN and PEFAC analyse the input speech using digital signal processing (DSP) operations whereas DNN-HMM and the proposed methods with a DNN regression model (\emph{DNN-REG}) and with an RNN regression model (\emph{RNN-REG}) exploit machine learning (ML) to estimate the $F0$ contour of speech. The key features of the comparative methods are summarized in Table \ref{tab:methods}.
\begin{table}[htbp]
	\begin{center}
		\caption{\it Key features of each $F0$ estimation method used in tests. AC, DP, SDF, AM and MF are abbreviations for autocorrelation, dynamic programming, squared difference function, aperiodicity measure and matching filter.}
		\label{tab:methods}
        \begin{tabular}{l|c|c|c}
            \multirow{2}[1]{*}{Method} & \multirow{2}[1]{*}{Approach} & Signal & \multirow{2}[1]{*}{Analysis} \\
            & & Domain &  \bigstrut[b]\\
            \hline
            RASP & \multirow{3}[2]{*}{DSP} & Time  & AC + DP \bigstrut[t]\\
            YIN & & Time  & SDF + AM \\
            PEFAC & & Log-Freq. & AC + MF + DP \bigstrut[b]\\
            \hline
            DNN-HMM & \multirow{3}[1]{*}{ML} & Log-Freq. & Classification \bigstrut[t]\\
            DNN-REG & & Freq. & Regression \\
            RNN-REG & & Freq. & Regression
        \end{tabular}
	\end{center}
\end{table}

Hyperparameters of the neural network models, i.e.\ DNN-HMM, DNN-REG and RNN-REG, are empirically selected by cross-validation tests with the CV dataset. The number of hidden layers are set equal to three with 1024 units each, and the mini-batch size is set to 200 frames. In DNN-REG and DNN-HMM the hidden layers are activated by ReLU function \cite{dahl13}. Random unit dropout (50 \%) and batch normalisation \cite{ioffe15} with momentum of 0.9 are applied during training. Hidden layers of RNN-REG are activated by $tanh$ function.

To capture temporal dynamics of input signals, seven previous frames and seven following frames are concatenated with the target frame and then input to the DNN in DNN-REG and DNN-HMM whereas fifteen consecutive time sequences centring the target frame input to the RNN in RNN-REG. The preceding hyper parameter settings are summarised in Table \ref{tab:setting}.
\begin{table}[htbp]
	\begin{center}
		\caption{\it Hyper parameter settings for DNN-HMM, DNN-REG and RNN-REG. (*: applied only to training)}
		\label{tab:setting}
		\begin{tabular}{l|c|c|c}
			Parameter & DNN-HMM & DNN-REG & RNN-REG\\
			\hline
			Output Layer & Classification& Regression & Regression\\
			\multicolumn{1}{r|}{\#units} & 68 & 1 & 1\\
			\multicolumn{1}{r|}{activation} & Softmax & Identity & Identity\\
			\hline
			Hidden Layer & Forward & Forward & RNN\\
			\multicolumn{1}{r|}{\#layers} & 3 & 3 & 3\\
			\multicolumn{1}{r|}{\#units} & 1024 & 1024 & 1024\\
			\multicolumn{1}{r|}{activation} & ReLU & ReLU & $\tanh$\\
			\multicolumn{1}{r|}{dropout} & $0.5^*$ & $0.5^*$ & No\\
			\multicolumn{1}{r|}{batch norm.} & Yes$^*$ & Yes$^*$ & No\\
			\hline
			Input & 1,005 dim & 7,695 dim & 513 dim\\
			\multicolumn{1}{r|}{mini-batch} & 200 & 200 & 200\\
			\multicolumn{1}{r|}{context} & 7 + 7 & 7 + 7 & 7 + 7
		\end{tabular}
	\end{center}
\end{table}

\subsection{Metrics of performance}
Performance of the $F0$ contour estimation methods %(DNN-REG and RNN-REG) to estimate the $F0$ contour of speech 
is evaluated using standard metrics used in $F0$ tracking literature: gross pitch error (GPE) rate and fine pitch error (FPE) \cite{rabiner76}. %and compered with RAPT, YIN, PEFAC and DNN-HMM.
GPE frames are voiced frames in which the error between the estimate of pitch period ($1/\hat{f}0$) and the ground truth ($1/f0$) is more than the period corresponding to 10 samples, i.e.\ 0.625 ms. Therefore, GPE rate is determined as
\begin{equation}
	\text{GPE rate} = \frac{N_{\text{GPE}}}{N_v}
\end{equation}
where $N_{\text{GPE}}$ and $N_v$ denote the number of GPE frames and voiced frames per utterance respectively.
FPE frames, in turn, are voiced frames excluding GPE frames. Mean of FPEs, $\mu_{\text{FPE}}$, represents the bias in $F0$ estimation whereas Standard deviation of FPEs, $\sigma_{\text{FPE}}$, measures the accuracy of estimation \cite{rabiner76}.
\begin{eqnarray}
	\mu_{\text{FPE}} &=& \frac{1}{N_{\text{FPE}}}\sum_{i=1}^{N_{\text{FPE}}}\epsilon_i\\
	\sigma_{\text{FPE}} &=& \sqrt{\frac{1}{N_{\text{FPE}}}\sum_{i=1}^{N_{\text{FPE}}}\left(\epsilon_i - \mu_{\text{FPE}}\right)^2}\\
	\epsilon_i &=& \left|\hat{f}_0^i - f_0^i\right|
\end{eqnarray}
where $\hat{f}_0^i$ and $f_0^i$ denote the estimate and grand truth of $F0$ respectively at the $i$-th frame in FPE frames while $N_{\text{FPE}}$ is the number of FPE frames.

\subsection{Results and discussion}
\begin{figure}[htb]
	\begin{center}
		\includegraphics[scale=0.73]{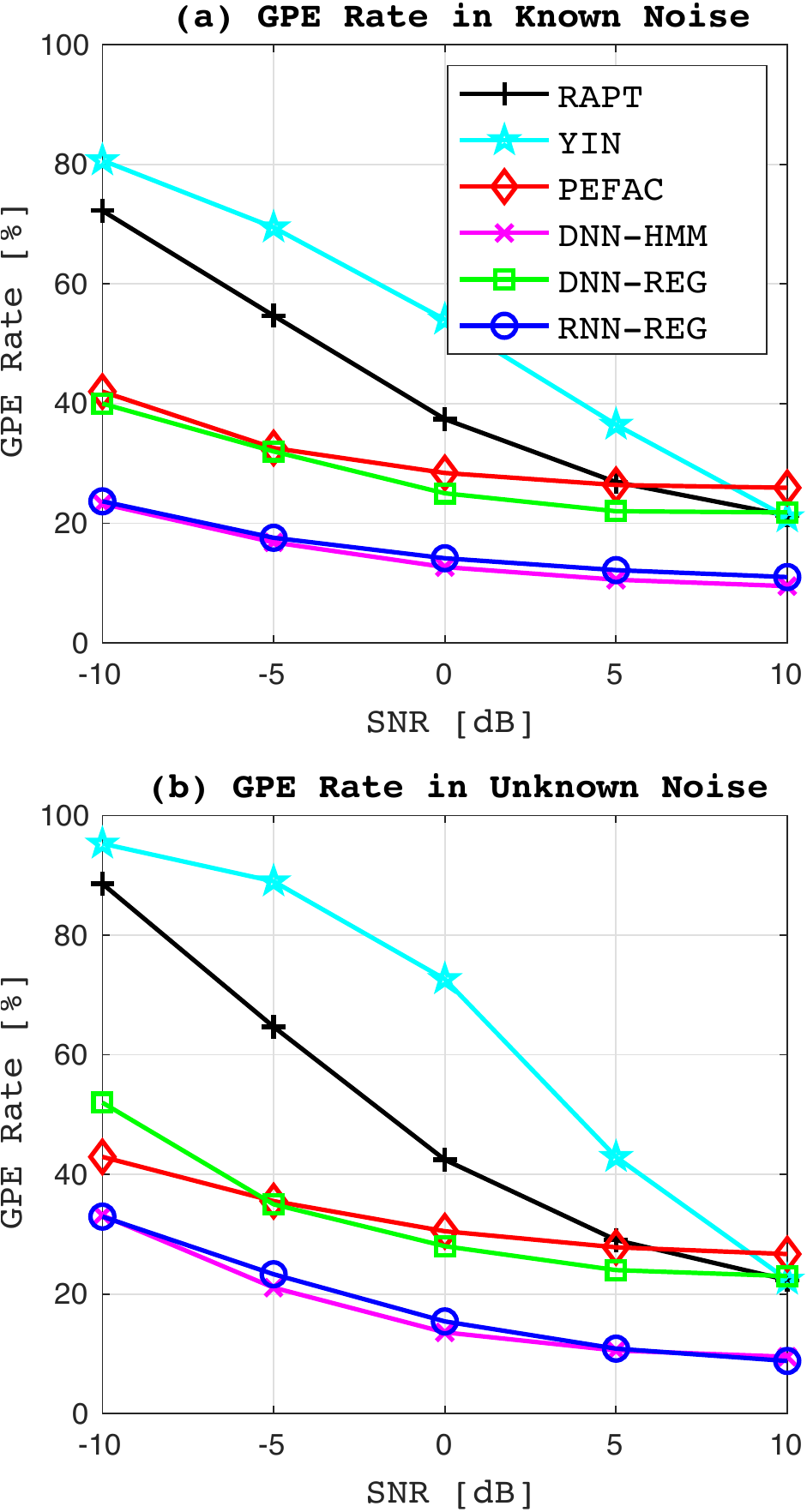}
		\caption{\it GPE rate of each method at SNRs of -10, -5, 0, 5, 10 dB in (a): the known noise condition including Babble, F16, Leopard, Machinegun, Volvo and White noise and (b): the unknown noise condition comprising Factory1 and Pink noise.}
		\label{fig:result_gross}
	\end{center}
\end{figure}

Figure \ref{fig:result_gross} (a) illustrates GPE rates of each method at different SNRs in the multi noise condition including Babble, F16, Leopard, Machinegun, Volvo and White noise, which are also shown during training, i.e.\ the known noise condition. (b) represents GPE rates in Factory1 and Pink noise as the unknown noise condition. 
RNN-REG shows the performance on almost same level as DNN-HMM in terms of GPE rate. They are superior to the other methods over the SNR range between -10 dB and 10 dB in both known and unknown noise conditions giving GPE rate of around 22 \% at -10 dB in known noise although it increases to 33 \% in unknown noise. PEFAC and DNN-REG also show noise-robustness as compared with YIN and RAPT but GPE rates are always from 10 to 20 percentage point higher than RNN-REG and DNN-HMM.

GPE frames are equivalent to failure in $F0$ estimation at voiced frames \cite{rabiner76}. In that sense, $F0$ estimation with YIN at SNRs below 10 dB, RAPT at less than 0 dB and PEFAC and DNN-REG at -10 dB and below are likely to have unreliable frames accounting for more than 40 \% of voiced frames. Conversely, RNN-REG and DNN-HMM keep estimation failure below 33 \% of voiced frames even at -10 dB in unknown noise. This brings substantial advantage in $F0$ contour estimation from noisy speech.

\begin{figure}[htb]
	\begin{center}
		\includegraphics[scale=0.73]{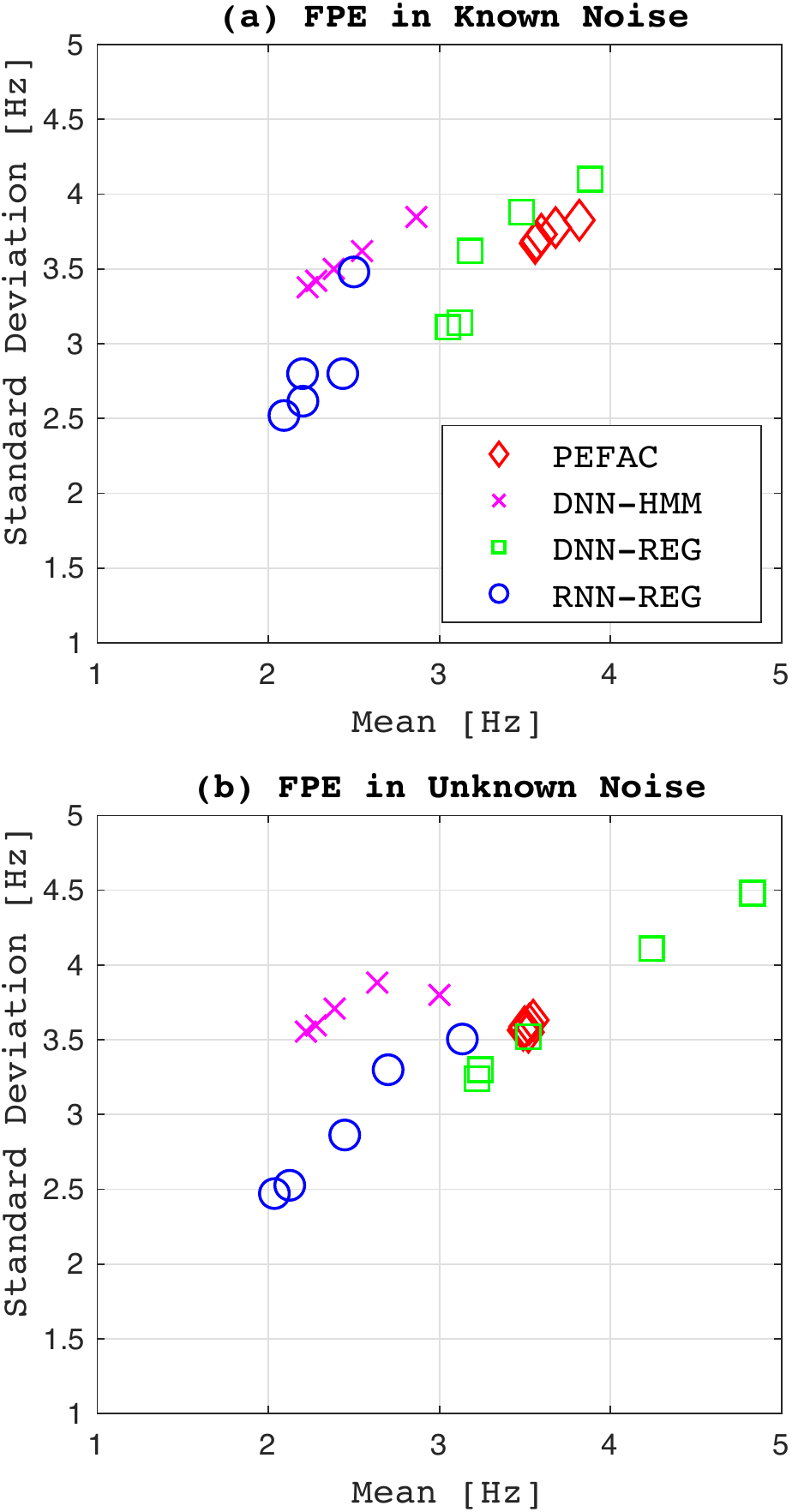}
		\caption{\it scatter plots of $\mu_{\text{FPE}}$ and $sigma_{\text{FPE}}$ at SNRs of -10, -5, 0, 5, 10 dB in (a): the known noise condition including Babble, F16, Leopard, Machinegun, Volvo and White noise and (b): the unknown noise condition comprising Factory1 and Pink noise.}
		\label{fig:result_fine}
	\end{center}
\end{figure}

Figure \ref{fig:result_fine} (a) and (b) illustrate the performance of PEFAC, HMM-DNN, DNN-REG and RNN-REG in terms of FPE at SNRs of -10, -5, 0, 5 and 10 dB in the known and unknown noise conditions respectively as scatter plots of $\mu_{\text{FPE}}$ and $\sigma_{\text{FPE}}$. YIN and RAPT are eliminated from this evaluation because sufficient amount of frames for FPE analysis are not brought by those methods in such noisy conditions.

Since $\mu_{\text{FPE}}$ represents the bias in $F0$ estimation while $\sigma_{\text{FPE}}$ is a measure of the accuracy in the estimation \cite{rabiner76}, RNN-REG performs best in terms of both bias and accuracy of estimation over the SNR range between -10 dB and 10 dB in the known and unknown noise conditions. Although PEFAC shows strong noise-robustness in both accuracy and bias, RNN-REG outperforms it by 31 \% in known noise and 24 \% in unknown noise according to the distance of their centroids. DNN-HMM performs slightly better than PEFAC but the performance against RNN-REG is lower by 17 \% in both known and unknown noise conditions. DNN-REG performs on the same level as PEFAC in known noise but it substantially loses noise-robustness in unknown noise and thus, the performance at SNRs of -5 dB and below in unknown noise is behind the other three methods.

In comparison between DNN-REG and DNN-HMM, the classification model in DNN-HMM performs better than the DNN regression model in DNN-REG in terms of GPE rate and FPE because the regression task to map noisy power spectra onto the exact $F0$ value is more difficult than the classification task to classify them into quantised frequencies. However, RNN regression improves the DNN regression by capturing temporal dynamics by optimising recurrent weights unlike DNNs augmenting the input with consecutive frames which produce a lot of poor-correlated connections into the network, e.g. a connection between the first bin in a past frame and the last bin in a future frame. Consequently, RNN regression accuracy outperforms the resolution of the quantised frequencies in the classification task.

\begin{figure}[htb]
	\begin{center}
		\includegraphics[scale=0.67]{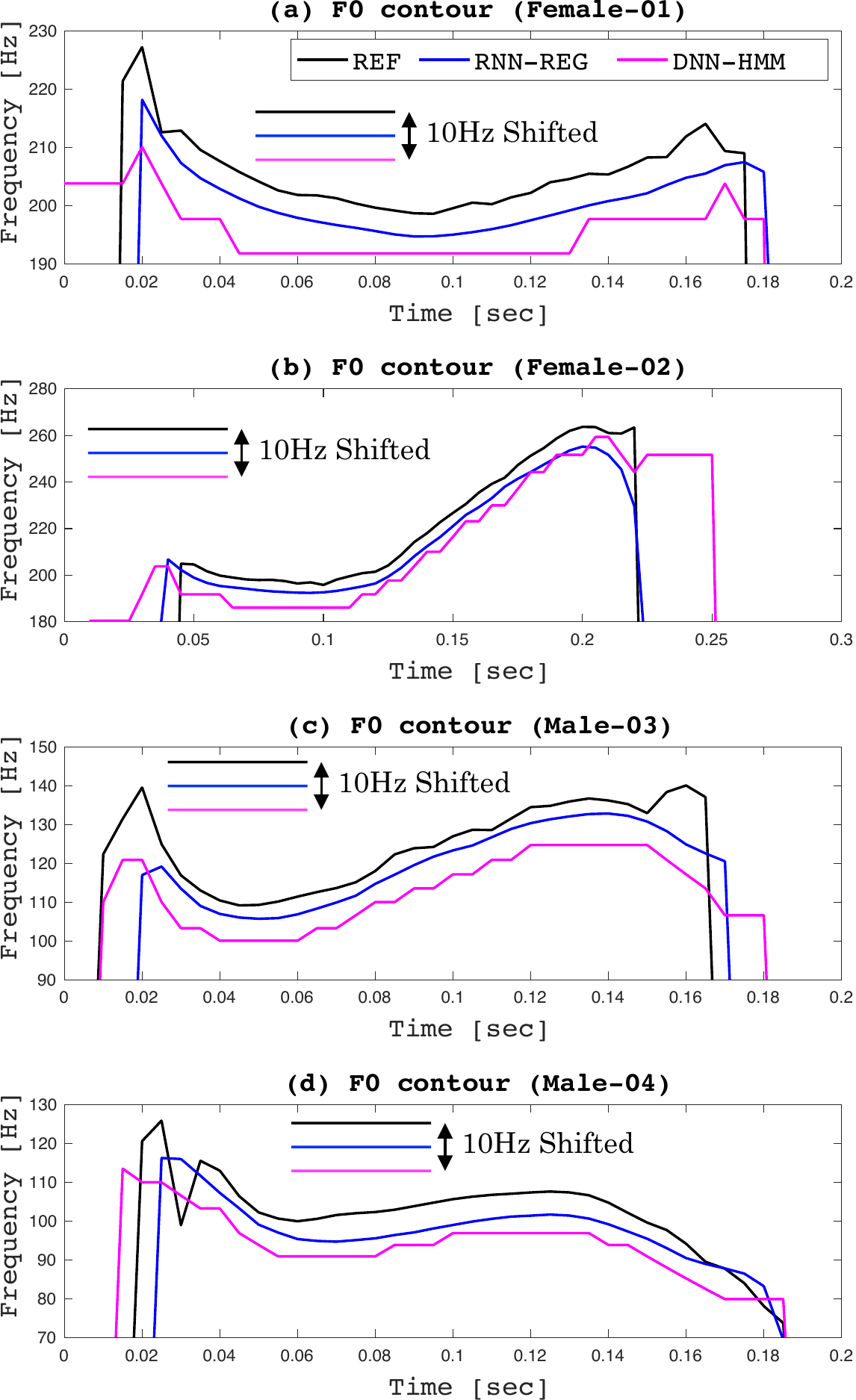}
		\caption{\it $F0$ contours of word `DARK" spoken by (a) Female speaker-01, (b) Female speaker-02, (c) Male speaker-03 and (d) male speaker-04. The $F0$ contours are estimated by RNN-REG and DNN-HMM and compared with the ground truth (REF).}
		\label{fig:contour}
	\end{center}
\end{figure}

Figure \ref{fig:contour} illustrates $F0$ contours of spoken word ``DARK" estimated by RNN-REG and DNN-HMM in a clean condition and they are compared with the ground truth (\emph{REF}). (a), (b), (c) and (d) show the $F0$ contours spoken by female speaker-01, female speaker-02, male speaker-03 and male speaker-04 respectively. Utterances of these four speakers are not included in the training dataset, i.e.\ they are unknown speakers.
The figures demonstrate the advantage of our RNN regression approach to $F0$ contour estimation over DNN-HMM representing the classification approach showing that the $F0$ contours estimated by RNN-REG is closer to the ground truth and more natural than DNN-HMM. Simultaneously, they clarify higher potential of the proposed method (RNN-REG) to track prosody of different speakers.

\section{Conclusion}
We addressed the problem of $F0$ contour estimation by using DNN and RNN-based regression techniques, with the aim of obtaining accurate $F0$ estimates with improved noise-robustness. While the DNN-based approach failed to provide accurate regression for the improvement, the RNN-based variant shows considerable achievement. Compared to PEFAC, one of the most noise-robust autocorrelation-based $F0$ trackers, the proposed method yielded a relative improvement exceeding 20\% in gross pitch error (GPE) rate at SNRs between -10 dB and +10 dB in unknown noise conditions. Furthermore, our RNN-based regression model outperformed a state-of-the-art, DNN-HMM-based $F0$ tracker, in terms of fine pitch error (FPE) by approximately 20 \% without substantially impacting GPE. 

Comparison of the estimated $F0$ contours of clean speech demonstrates an advantage of the proposed method over DNN-HMM approach in producing more natural $F0$ trajectories. This work focused solely on the $F0$ tracking itself, but our near-future plans involve integrating our proposal to applications such as voice conversion and prosody-based speaker and language recognition.

%Therefore, this work is expected to contribute to further development in speech and spoken language processing by begin applied to, for instance, prosody analysis, voice conversion and speech synthesis as well as speaker and language identification. 

% -------------------------
% ORIGINAL CONCLUSION HERE:
% -------------------------
%
% We propose an approach using an RNN-based regression model to $F0$ contour estimation in order to obtain more accurate and noise-robust performance than existing $F0$ trackers. The proposed RNN-based method brings improvement in GPE rate by more than 20 \% at SNR of -10 dB even in unknown noise conditions as compared with PEFAC, which is one of the most noise-robust autocorrelation-based $F0$ tracker. Moreover, it outperforms a state-of-the-art DNN-HMM-based $F0$ tracker in terms of FPE by approximately 20 \% while keeping the same level of performance in terms of GPE rate. Comparison of estimates of $F0$ contours in clean condition also shows an advantage of the proposed method over DNN-HMM approach showing more natural trajectory of the $F0$ contours. Therefore, this work is expected to contribute to further development in speech and spoken language processing by begin applied to, for instance, prosody analysis, voice conversion and speech synthesis as well as speaker and language identification. 

\section{Acknowledgement}
This work was supported in part by Academy of Finland (Proj. No. 309629). The authors wish to acknowledge CSC - IT Center for Science, Finland, for computational resources.

\bibliographystyle{IEEEbib}
\bibliography{odyssey2018.bib}

\end{document}